\documentclass[aps,prx,twocolumn,superscriptaddress,noeprint,longbibliography, nofootinbib]{revtex4-1}

\usepackage{graphicx}
\usepackage{bm}
\usepackage{physics}
\usepackage{xcolor}
\usepackage{enumitem}
\usepackage{amsmath, amssymb}
\usepackage[normalem]{ulem}
\usepackage{natbib}
\usepackage{bbm}
\usepackage{comment}

\usepackage[colorlinks=true]{hyperref}
\hypersetup{citecolor = blue}

\newcommand{\barphi}{\bar{\phi}}
\newcommand{\conjphi}{\phi^*}
\newcommand{\expterm}{\eta}
\newcommand{\lrpow}{\kappa}
\newcommand{\pann}{p_{\text{ann}}}
\newcommand{\pbr}{p_{\text{br}}}
\newcommand{\pc}{p_c}
\newcommand{\decaypow}{ {\alpha} }
\newcommand{\decaypowc}{ {\alpha_c} }
\newcommand{\decaypowDP}{ {\alpha_c^{\text{DP}}} }
\newcommand{\decaypowPC}{ {\alpha_c^{\text{PC}}} }
\newcommand{\numneigh}{\binom{2d}{2}}
\renewcommand{\b}{\hat{b}}
\DeclareMathOperator*{\Motimes}{\text{\raisebox{0.25ex}{\scalebox{0.8}{$\bigotimes$}}}}

\newcommand{\ham}{\hat{H}}
\newcommand{\unit}{\mathbbm{1}}

\begin{document}

\title{Absorbing state transitions with long-range annihilation}

\author{Nicholas O'Dea}
\affiliation{Department of Physics, Stanford University, Stanford, CA 94305, USA}
\affiliation{Princeton Center for Theoretical Science, Princeton University, Princeton, NJ 08544, USA}
\affiliation{Princeton Quantum Initiative, Princeton University, Princeton, NJ 08544, USA}
\author{Sayak Bhattacharjee}
\affiliation{Department of Physics, Stanford University, Stanford, CA 94305, USA}
\author{Sarang Gopalakrishnan}
\affiliation{Department of Electrical and Computer Engineering, Princeton University, Princeton, NJ 08544, USA}
\author{Vedika Khemani}
\affiliation{Department of Physics, Stanford University, Stanford, CA 94305, USA}

\begin{abstract}
We introduce a family of classical stochastic processes describing diffusive particles undergoing branching and \emph{long-range annihilation} in the presence of a parity constraint. The probability for a pair-annihilation event decays as a power-law in the distance between particles, with a tunable exponent. Such long-range processes arise naturally in various classical settings, and increasingly play a role in the study of quantum dynamics: for example, quantum state preparation or error correction processes aim to prepare ordered ground states, which requires removing point-like excitations in pairs via \emph{non-local} feedback operations conditioned on a \emph{global} set of measurement outcomes. We analytically and numerically describe features of absorbing phases and phase transitions in this family of classical models as pairwise annihilation is performed at larger and larger distances. Notably, we find that the two canonical absorbing-state universality classes -- directed-percolation and parity-conserving -- are endpoints of a new line of universality classes with continuously varying critical exponents. These new classes follow from an interplay of long-range interactions and symmetries: decreasing the long-range exponent effectively weakens the local effects of global symmetry constraints. This competition significantly broadens known non-equilibrium phases and phase transitions.

\end{abstract}

\maketitle

\noindent \textbf{\textit{Introduction--}}
Absorbing state transitions from a fluctuating active phase to an inactive absorbing state are widespread in non-equilibrium classical settings, including turbulence, catalysis, and epidemic spreading; however, despite the rich variety of systems modeled by these phase transitions, there are few known universality classes~\cite{hinrichsen2008nonequilibrium}. In this work, we demonstrate that competition between global symmetries and long-range annihilation processes substantially broadens the classes of known absorbing state phases and phase transitions.

Long-range interactions are natural in many models with absorbing state transitions, from chemical reactions involving charged particles, to predator-prey models incorporating movement patterns of different species (see Ref.~\cite{hinrichsen2007longrange} for a review discussing long-range hopping and branching, and Refs.~\cite{suchanpark2020first, suchanpark2020second} for recent work on biased hopping). However, models with long-range annihilation have largely been neglected in the classical absorbing state literature. An exception is Ref.~\cite{park1999field} by Park and Deem, which, however, does not consider branching processes and does not study the phase transition from the absorbing phase to the active phase.

\begin{figure}[t]
\includegraphics[width=\linewidth]{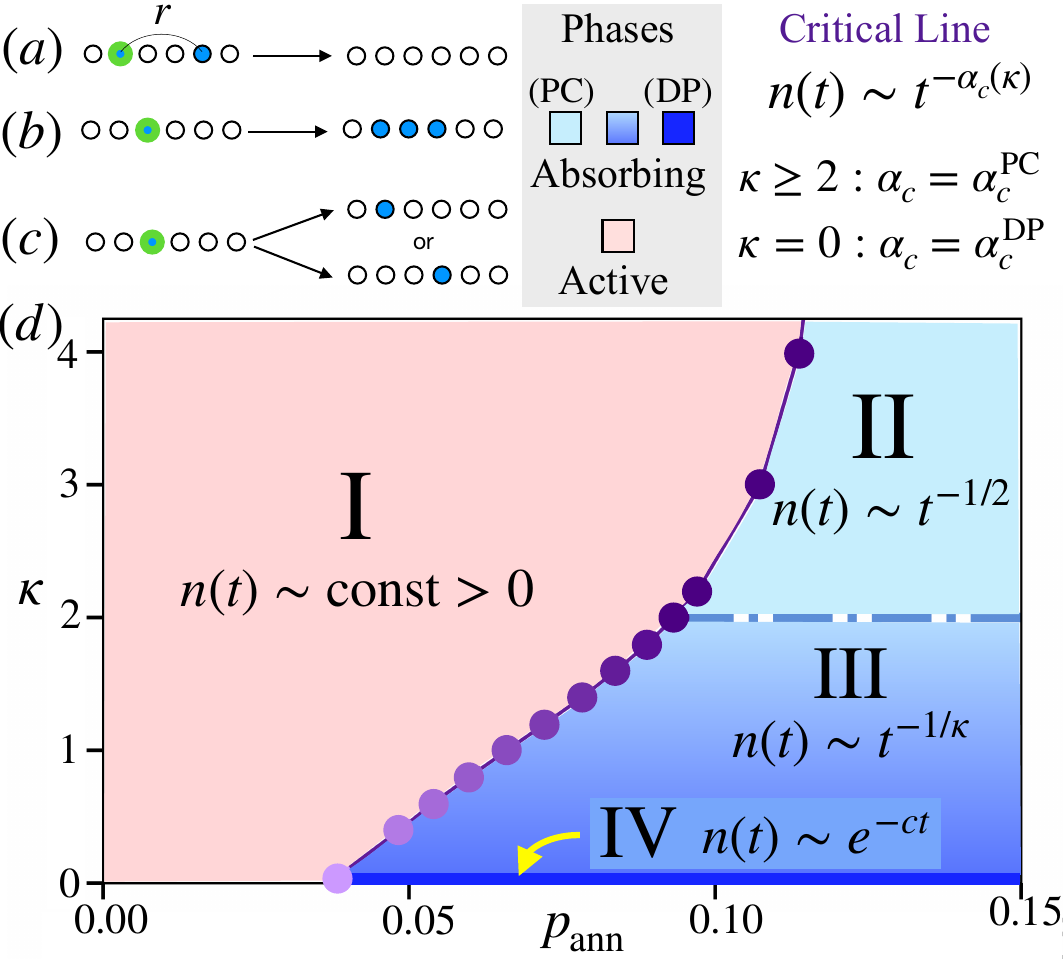}
 \caption{Rules for (a) long-range annihilation controlled by $p_{\rm ann}$ and $\kappa$, (b) branching set by $p_{\rm br}=.2$, and (c) hopping. (d) Phase diagram depicting active (I) and absorbing (II-IV) phases, with a line of critical points with exponents interpolating from the parity-conserving (PC) universality class for $\lrpow\geq 2$ to the directed percolation (DP) universality class at $\lrpow=0$. The absorbing phases II, III, and IV correspond to distinct scaling regimes of the decay of the particle density and are not separated by phase transitions; the horizontal dashed line between II and III at $\kappa=2$ is the ceiling for $\kappa$-dependent decay exponents. In the absorbing phases and along the transition line, we schematically use darker colors to denote larger decay exponents; while $\alpha$ increases with smaller $\kappa$ (Fig.~\ref{fig:absorbing_decay}), $\alpha_c$ decreases with smaller $\kappa$ (Fig.~\ref{fig:transition_decay}).   }
\label{fig:rules_and_pd}
\end{figure}

To address this hole in the literature, we introduce a stochastic reaction-diffusion process describing particles undergoing branching and long-range annihilation in the presence of a parity constraint which enforces that particles are created and annihilated in pairs (Fig.~\ref{fig:rules_and_pd}(a)-(c)). By tuning a long-range exponent $\kappa$, we analytically and numerically probe how the features of the absorbing phase and the transition change as pairwise annihilation is successfully performed at larger and larger distances.

There is a competition between this long-range annihilation and the parity symmetry. Heuristically, consider a long-range annihilation process that destroys a pair of far-separated particles. To an observer in the local environment of each particle, the pairwise annihilation step appears to remove that particle without regard to the parity symmetry. Only upon viewing the whole system does the global parity constraint become manifest. 

This competition leads to striking consequences. Two paradigmatic universality classes of absorbing state transitions are directed percolation (DP) and parity-conserving (PC), respectively corresponding to local models without and with a parity constraint. Having a parity constraint or not seems like a binary question, making DP and PC discrete and ostensibly unconnected universality classes. However, our model uncovers a line of universality classes with \textit{exponents that continuously interpolate between DP and PC} despite a global parity constraint (Fig.~\ref{fig:rules_and_pd}(d)), greatly enriching the set of known critical phenomena in non-equilibrium phase transitions. 

In long-range equilibrium and non-equilibrium systems, continuously varying exponents are usually associated with an approach to mean-field physics under sufficiently-long-range interactions (e.g. section 3 of Ref.~\cite{hinrichsen2007longrange}). However, we find in numerics on a chain that the transition in the long-range limit corresponds to $(1+1)d$ DP, \textit{not} mean-field DP. Our numerics are for one dimension, but we discuss scaling arguments that should be valid in the absorbing phase in all dimensions. We probe the transition numerically, and we outline a field theory program to understand its properties. We derive the field theory describing our model in Appendix~\ref{app:fieldtheory} and discuss some of its interesting features in the main text. Though field theories usually have actions that are polynomial in the fields, we have a nontrivial exponential term in our action. We show that the resulting mean-field theory in the absorbing phase crucially relies on this exponential term and matches the predictions of our simple scaling arguments in $d \geq 2$.

\noindent \textbf{\textit{Model and phase diagram--}} 
We study a classical stochastic process describing a system of particles undergoing diffusion, branching and annihilation subject to a parity constraint which conserves the number of particles modulo two. Our model consists of a mixture of deterministic and stochastic steps, depicted in Fig.~\ref{fig:rules_and_pd}(a)-(c) for a system in one spatial dimension. Each time step begins by selecting a particle at random, illustrated by the green outline. With probability $\pann$, we continue to the annihilation stage in (a), where the particle's distance $r$ to the nearest excitation is calculated and with probability $1/r^\lrpow$ the two particles are annihilated. Thus, $\pann$ sets the rate at which pairing is attempted, while $\lrpow$ controls the nonlocality of the annihilation. Because we only attempt to annihilate a particle with the nearest particle rather than attempting pairwise annihilation with all other particles, the thermodynamic limit is already well-defined \textit{without} any system-size dependent rescaling of the attempted annihilation rate (e.g. as in Ref.~\cite{botzung2021kac}), even for the extreme long-range case of $\kappa=0$.

If, with probability $1-\pann$, annihilation is not selected, then (b) with probability $\pbr$
the particle branches (produces two offspring on two different adjacent sites) or (c) with probability $1-\pbr$, it attempts to hop to a neighboring site. A hard-core constraint is enforced by aborting the attempt to branch or hop if more than one particle would occupy the same site. At the end of each time step, in a system with $N$ particles, the time increments by $1/N$ to model these processes occurring in parallel; the more particles, the more of these processes happen during a given time interval. Note that a system with no particles is an absorbing state without dynamics; this state can be reached but never left.  In our numerics, we work in one spatial dimension and fix $\pbr = .2$, while we explore the phase diagram shown in Fig.~\ref{fig:rules_and_pd}d) with parameters $\pann$ and $\lrpow$. Throughout, we consider initial states with an even number of particles so that in principle the absorbing state can be reached; in particular, we choose even-length chains and initialize in the fully occupied state. 

When the annihilation rate $\pann$ is low, the density of particles $n(t)$ is asymptotically constant up to a time growing exponentially in system size. This ``active" phase is denoted by ``I" in Fig.~\ref{fig:rules_and_pd}d). When $\pann$ is sufficiently large, the density of particles decays to zero in a time at most polynomial in system size. The functional form of the decay rate $n(t) \sim t^{-\decaypow}$ in these ``absorbing" phases is set by $\lrpow$ (see phases II, III, IV in Fig.~\ref{fig:rules_and_pd}d)). Between the active and absorbing phases is a critical line of absorbing state transitions where the density decays as $n(t) \sim {t^{-\decaypowc}}$; we denote the critical decay exponent $\decaypowc$ to differentiate it from the absorbing phase decay exponent of $\decaypow$. We summarize our results for the phase diagram in Fig.~\ref{fig:rules_and_pd}d), and provide a detailed exposition below.

\noindent \textbf{\textit{Absorbing phase analytics and numerics--}}
The properties of the absorbing phases are controlled by the pure annihilation fixed point and hence can be understood through scaling arguments in the limit of no branching. 
Though our main focus in this work is the case of $d=1$, we also consider the characteristics of the absorbing phase in higher dimensions. 

We note that section 3.1 of Ref.~\cite{park1999field} discusses a related model with both long-range annihilation and diffusion, and in particular notes an analogous phase to our PC absorbing phase II in Fig.~\ref{fig:rules_and_pd}. However, that section restricts to $\kappa > \min(2,d)$ (and the paper restricts to $\kappa > d$ more generally), and so does not obtain our regimes III and IV. 
Furthermore, the existence and properties of the absorbing state \textit{transition} (our next section) require branching and are not considered in Ref.~\cite{park1999field}.
Our model also has an emergent cutoff on the range of the interaction. By only pairing a particle to the particle closest to it, we avoid diverging annihilation rates in the thermodynamic limit even when the nonlocality of the annihilation is maximal. Together, branching and the natural cutoff give us access to a new regime of the phase diagram; this otherwise hidden part of the phase diagram contains the novel phase transitions that interpolate all the way between the PC and DP universality classes.

First, consider the short-range limit of $\lrpow = \infty$ where particles must be neighboring in order for annihilation to proceed. Annihilation is then diffusion-limited, as particles must effectively collide in order to annihilate. In this limit, there is a standard set of arguments for the decay of $n(t)$ in terms of the number of unique sites visited by a random walker in a time $t$. The number of unique sites visited controls the number of particles a given particle can encounter as it diffuses and hence the number of annihilation events (see chapter 13 of Ref.~\cite{shlomo2000textbook} for more details, and Ref.~\cite{peliti1986pureannihilation} for alternative arguments using field theory).
In particular, in $d$ dimensions
\begin{equation}\label{eq:difflimit}
    n(t) \sim \begin{cases} {t^{-d/2}} & d < 2 
    \\ t^{-1}{\log(t)} & d = 2
    \\ {t^{-1}} & d > 2
    \end{cases}
\end{equation}

On the other hand, for $\lrpow < \infty$, there will be a steady background decay of particles from the long-range annihilation. Suppose $\lrpow$ is small so that the particle annihilation rate is dominated by events where the particles are not necessarily close. In $d$ dimensions, the typical interparticle distance will scale on the order of $r\sim {n(t)^{-1/d}}$ and hence the typical probability for a pairing to result in annihilation will scale as $n(t)^{{\lrpow}/{d}}$. The number of annihilation events per unit time will scale as the number of particles times this probability, giving an effective rate equation $\dot{n}(t) \sim n(t)^{1+{\lrpow}/{d}}$. Solving this equation gives
\begin{equation}\label{eq:lrlimit}
    n(t) \sim \begin{cases} {t^{-d/\lrpow}} & \lrpow >0
    \\ e^{-c t} & \lrpow = 0 \end{cases}
\end{equation}
With diffusion turned off, this result (for $\kappa>0$) has been rigorously proven in Ref.~\cite{burlatsky1990asymptotics}, and further discussed in a renormalization group context in Ref.~\cite{park1999field} (for $\kappa > d$).

We now make the simple assumption that the process (short-range annihilation induced by diffusion vs. long-range annihilation between distant particles) which leads to a faster decay is the one that (asymptotically) dominates the dynamics. Comparing Eqs.~\ref{eq:difflimit} and ~\ref{eq:lrlimit}, we get in $d=1$:
\begin{equation}\label{eq:competition}
    n(t)  \sim \begin{cases} {t^{-1/2}}& \lrpow \geq 2 \,\,\,\,\,\,\,\, \,\,\,\,\,\,\,\text{(II)}
    \\ {t^{-1/\lrpow}} & 0 < \lrpow \leq 2 \,\,\,\, \text{(III)}
    \\ e^{-c t} & \lrpow = 0 \,\,\,\, \,\,\,\,\,\,\,\,\,\,\,\text{(IV)}
    \end{cases}
\end{equation}
The parenthetical labels (II/III/IV) are shown in the corresponding regions of the phase diagram in Fig.~\ref{fig:rules_and_pd}d). 

\begin{figure}[t]
\includegraphics[width=\linewidth]{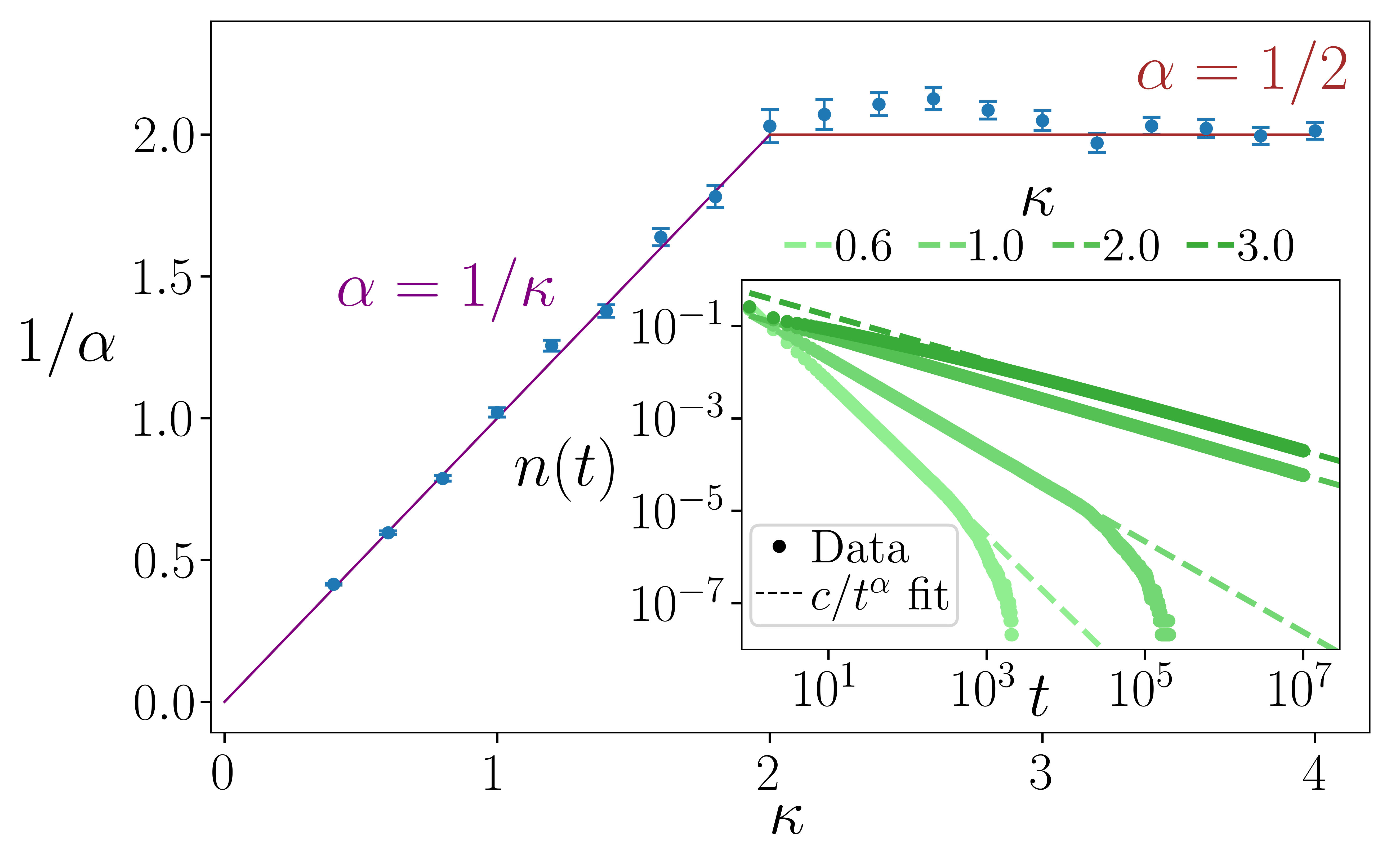}
\caption{Dynamics deep in the absorbing phase ($\pann = .8$) for $L=96000$. Between $800$ - $4000$ samples were averaged to construct $n(t)$ at each $\kappa$ value probed. In the absorbing phase, the density decays as $n(t) \sim t^{-\decaypow}$, and this figure compares the prediction in Eq.~\ref{eq:competition} to $\decaypow$ extracted from the dynamics. Define $t_f$ as $10^7$ or the time when $n(t)=10^{-4}$, whichever is smaller. $\decaypow$ is extracted via linear fits of $\log(n(t))$ against $\log(t)$ for times between $t_f/2$ and $t_f$, and error bars are standard deviations from $100$ bootstrap resamples of $n(t)$. The inset shows $n(t)$ against $t$ along with fits (dotted lines) for a handful of $\lrpow$ values.
}
\label{fig:absorbing_decay}
\end{figure}

We numerically confirm this picture in Fig.~\ref{fig:absorbing_decay} via linear fits of $\log(n(t))$ vs $\log(t)$. There is good agreement except for a small, systematic discrepancy around $\lrpow \approx 2.6$. As we discuss in Appendix~\ref{app:subleading}, We believe this discrepancy is caused by asymptotically sub-leading terms from the background long-range decay from $\lrpow$ in this otherwise diffusion-limited regime. For example, $\lrpow=3$ in the inset shows an ultimately-transient slower decay at early times. 

Similarly, for $d \geq 2$,
\begin{equation}\label{eq:competition_bigd}
    n(t) \sim \begin{cases} {t}^{-1} & \lrpow \geq d
    \\ t^{-d/\lrpow} & 0 < \lrpow \leq d
    \\ e^{-c t} & \lrpow = 0
    \end{cases}
\end{equation}
We save the numerical verification of this prediction in $d \geq 2$ for future work. To summarize, for $\lrpow > \max(2,d)$, the model is effectively short-ranged and the absorbing phase is diffusion-limited, while for $\lrpow < \max(2,d)$, absorption occurs  faster and is dominated by long-range annihilation. In particular, in $d=1$, the decay in the absorbing phases at $\lrpow \geq 2$ and $\lrpow=0$ are consistent with the PC and DP absorbing phases of $n(t) \sim t^{-1/2}$ and $n(t) \sim e^{-ct}$ respectively.

We can interpret these results in terms of the parity constraint. Conventionally, 
the DP/ PC universality classes are distinguished by the absence/ presence of a discrete parity symmetry.  In our model, the \emph{global} parity constraint exactly holds, but it no longer has consequences locally when $\kappa=0$. In particular, looking at just a local subsystem, long-range annihilation will violate the parity of the number of particles in the subsystem even as it preserves the global parity. This allows a model with a global parity constraint to be described by symmetry-free DP rather than PC physics. Thus we expect that the $\kappa = 0$ model is described by DP, including at the phase transition. We confirm this numerically in the next section.

\begin{figure}[t]
\includegraphics[width=\linewidth]{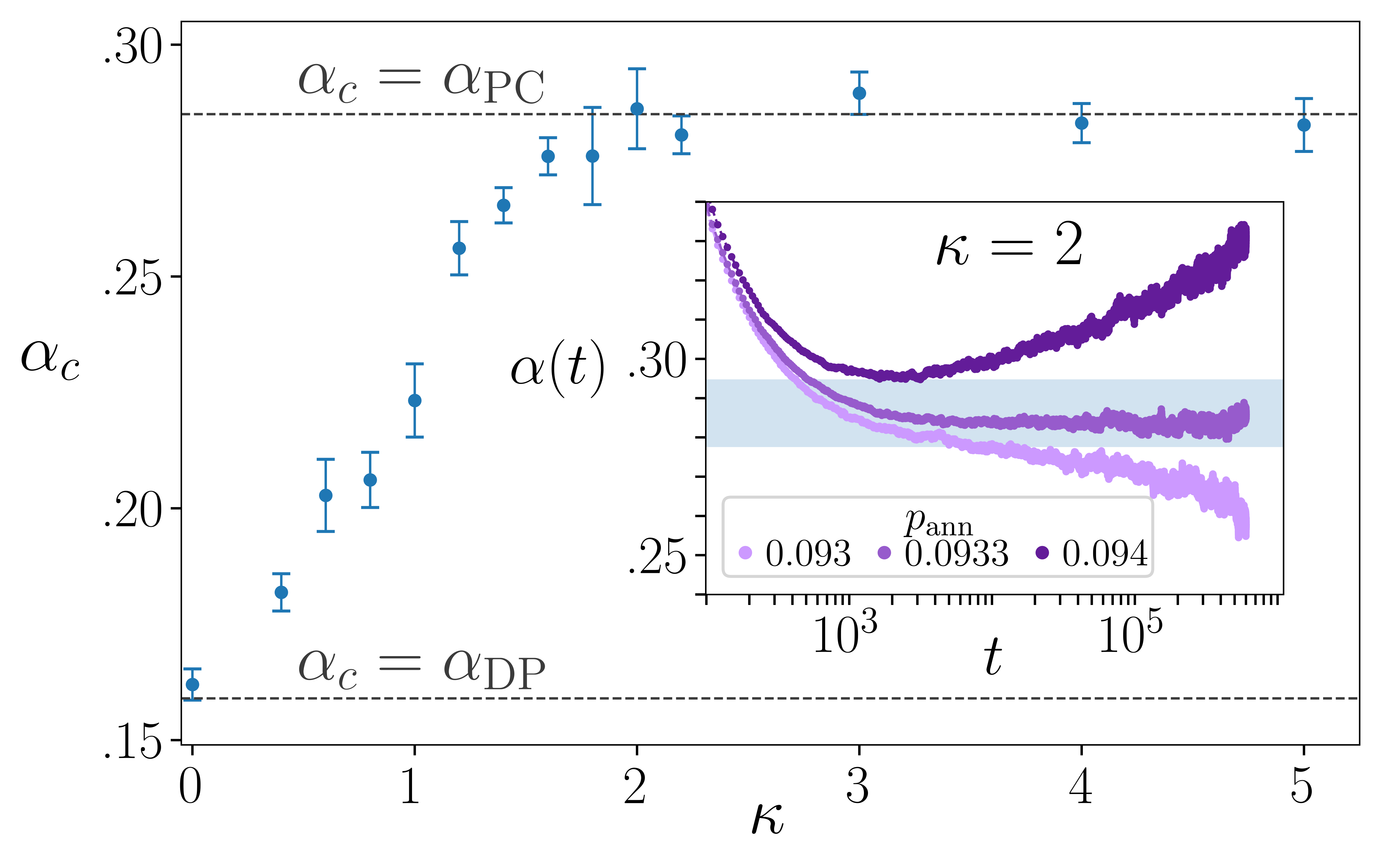}
\caption{Decay exponent $\decaypowc$ at the transition for $L=96000$. Between $800$ - $4000$ samples were averaged to construct $n(t)$ at each $\kappa$ and $p$ value probed. The transition location $\pc$ is determined by the separatrix. Inset shows an example for $\lrpow = 2$; shaded blue region is the numerically extracted $\decaypowc(\lrpow = 2)$ up to one standard deviation. Standard deviations are estimated with 100 bootstrap resamples. $\decaypowc$ at the separatrix is plotted in the main panel, and compared to $\decaypowDP$ and $\decaypowPC$. Note the different $y$-axis as compared to Fig.~\ref{fig:absorbing_decay}. }
\label{fig:transition_decay}
\end{figure}

\noindent \textbf{\textit{Transition numerics--}}
In Fig.~\ref{fig:transition_decay}, we find that the decay exponent $\decaypowc$ is consistent with DP's $\decaypowDP \approx .159$ at $\lrpow=0$ and PC's $\decaypowPC \approx .285$ at $\lrpow \geq 2$, and the exponents for $0 < \lrpow < 2$ appear to monotonically interpolate between the two, indicating a line of critical fixed points with continually varying exponents.  From our numerics, we cannot rule out that the critical exponents at the transition begin to change from $\decaypowPC$ at a $\lrpow_c$ slightly different from $2$. However, our numerics are consistent with the idea that the $\lrpow$ at which the phase behavior changes ($\lrpow = 2$ from the previous section) is the same $\lrpow$ at which the critical behavior changes. 
In the inset of Fig.~\ref{fig:transition_decay} we show an example of our method of extracting $\pc$ and $\decaypowc$ from the dynamics. At large system sizes and for $\pann$ close to $\pc$, the dynamics will appear to follow the algebraic decay of the critical model. However, when $\pann$ is detuned to be above or below the transition, the dynamics at late times will respectively drift to the different asymptotic behaviors of the absorbing and active phases. By calculating a ``running" $\decaypow(t) \equiv \log_{10}[n_d(t) / n_d(10t)]$, we see a horizontal separatrix between the absorbing and active phases; the $\pann$ and $\decaypow$ (after finite-time transients) of the separatrix give our estimates for $\pc$ and $\decaypowc$. 
See Ref.~\cite{mendonca2011stavskaya} for a succinct discussion of this and other numerical probes of absorbing state critical behavior and chapter 4 of Ref.~\cite{hinrichsen2008nonequilibrium} for a general discussion. Note that both DP and PC have large dynamical critical exponents (see Appendix~\ref{app:dynamical} for the interpolating universality classes), meaning that systematic late-time finite-size effects controlled by ${t}/{L^z}$ should only appear at much later times than we probe.

In the following section and in Appendix~\ref{app:fieldtheory}, we briefly outline a field theory program to analytically estimate the critical exponents and critical dimensions of these newly uncovered universality classes. 

\noindent \textbf{\textit{Field theory action--}}
Cardy and Täuber gave a systematic field theory treatment of critical phenomena and phase behavior in models with local, pairwise annihilation processes~\cite{cardytauber1998field} (see also the review~\cite{tauber2005review}). The starting point is the derivation of the field theory directly from the stochastic process~\cite{doi1976path, peliti1985path} by treating the real-time stochastic master equation as an imaginary-time Schrodinger equation. The new degrees of freedom are complex fields $\phi(x,t)$ and $\barphi(x,t)$; statistical expectation values of $n(x,t)$ can be computed via expectation values of $\phi(x,t)$ in the field theory (e.g. $\mathbb{E}[n(x,t)] = \langle \phi(x,t)\rangle$, $\mathbb{E}[n^2(x,t)] = \langle \phi^2(x,t) + \phi(x,t)\rangle$). 

Approximations like derivative expansions and the continuum limit for field theories near criticality simplify the field theory while maintaining the universality of the phase transition and phases, allowing for renormalization group calculations of critical exponents. However, the field theory keeps a few peculiar signatures of its origins as a classical stochastic process: conservation of probability ultimately forces the action to vanish whenever $\barphi = 1$, whatever the value of $\phi$; processes changing the number of particles make the theory explicitly non-Hermitian, which manifests as terms in the action carrying an unbalanced number of $\phi$ and $\barphi$. It is useful to identify $\phi(x,t)$ with $n(x,t)$ at the mean-field level and to build intuition for $\phi(x,t)$ from $n(x,t)$, but this identification cannot be made rigorously beyond the mean-field level~\cite{cardytauber1998field,tauber2005review}.

Using a method pioneered in~\cite{wijland2000hardcore}, we extend their results to derive a field theory for our nonlocal annihilation process in Appendix~\ref{app:fieldtheory}. Our action density $\mathcal{L}$ includes standard diffusion $\barphi(\partial_t-D \nabla^2)\phi$ and branching $B(1-{\barphi}^2)\barphi\phi$ terms, but it also contains an annihilation term that is exponentially suppressed in particle density:
\begin{equation}
\begin{split}
\mathcal{L(\mathbf{x}},t) & \ni  A \int \frac{d^d\mathbf{y}}{|\mathbf{x} - \mathbf{y}|^\lrpow}  \bigg[  \left(1-\barphi(\mathbf{x},t)\barphi(\mathbf{y},t)\right)\phi(\mathbf{x},t)\phi(\mathbf{y},t) \\& \textrm{exp}\left(-\int_{r=0}^{r=|\mathbf{x}-\mathbf{y}|} d^d \mathbf{r}\:   \barphi(\mathbf{x}+\mathbf{r},t)\phi(\mathbf{x}+\mathbf{r},t) \right) \bigg]
\end{split}
\end{equation}
The exponential suppression comes directly from the constraint that a particle can only be paired with the closest particle; heuristically, the exponential term is accounting for the low likelihood that two far away particles have no particles in between them. We believe this exponential factor plays a fundamental role in both the critical and absorbing phase behavior at small $\lrpow$; as we show in Appendix~\ref{app:EL_min} it plays a crucial role in mean-field theory.

\noindent \textbf{\textit{Discussion--}}
Our work mapped out the phase diagram of diffusive particles undergoing branching and long-range annihilation, and we uncovered a new line of universality classes connecting the naively unrelated DP and PC classes. We gave a heuristic interpretation of our results in terms of a competition between symmetries and long-range processes. While our model always has a global parity symmetry, long-range annihilation means this global parity constraint has weaker effects locally. Strikingly, this weakening occurs continuously as a function of the long-range exponent -- we found universality classes that are effectively intermediate between those with and without a symmetry. 

Long-range annihilation should allow similar interpolations for larger symmetry groups. Absorbing state transitions with larger symmetry groups (e.g. $\mathbb{Z}_2 \times \mathbb{Z}_2$) have not been investigated thoroughly, in part because the absorbing state transition is destroyed (these systems tend to be active for any nonzero branching~\cite{cardytauber1998field}); we expect that long-range annihilation should be able to discover nontrivial transitions in these models that interpolate to DP as an appropriate long-range exponent $\kappa \to 0$. 

In our analytics and numerics, we characterized the phases and phase transition by how the density of the particles decays with time starting from a completely occupied state, and we extend this analysis in Appendix~\ref{app:dynamical}.  Other probes of critical exponents include the survival time and growth of small seeds of particles in otherwise empty systems; we leave these complementary probes to future work.

Finally, we note that our study was inspired by considering quantum error correction processes which often require completely non-local feedback operations ($\kappa =0$) in various canonical cases. Despite this broad similarity, there are also crucial differences between descriptions of error correction and absorbing state transitions (Appendix~\ref{app:comp_qec}), and it would be interesting to explore how the different universality classes of absorbing state transitions as a function of $\kappa$ may inform error correction protocols with varying degrees of non-locality in feedback.

\noindent \emph{Acknowledgements--} We are grateful to Su-Chan Park for useful discussions about scaling arguments and mean field theory,  Alan Morningstar for prior collaboration on related work, and Pavel Nosov and Samuel Alipour-fard for discussions about field theory.
This work was supported by the Office of Naval Research Young Investigator Program (ONR YIP) under Award Number N00014-24-1-2098 (S.B and V.K.). 
V.K. also acknowledges support from the Alfred P. Sloan Foundation through a Sloan Research Fellowship and the Packard Foundation through a Packard Fellowship in Science and Engineering. N.O.D. acknowledges support
from the ARCS Foundation for ARCS Scholar funding and from the Princeton Center for Theoretical Science and the Princeton
Quantum Initiative at Princeton University.
Numerical simulations were performed on Stanford Research Computing Center's Sherlock cluster. Formulating the models was supported in part by the US Department of Energy Co-design Center for Quantum Advantage (C2QA)
under contract number DE-SC0012704 (S.G.).

\bibliography{main}

\onecolumngrid

\section*{Appendices}

We walk through the derivation of the field theory in Appendix~\ref{app:fieldtheory} with a mean field analysis in Appendix~\ref{app:EL_min}. In Appendix~\ref{app:subleading}, we discuss subleading terms in the absorbing phase density decay that lead to a small, systematic shift in the fit for the decay exponent. In Appendix~\ref{app:dynamical}, we numerically estimate the dynamical exponent $z_c$ as a function of $\kappa$ at the absorbing state phase transition. In particular, it is consistent with our story of the phase transition and shows qualitatively similar features to $\alpha_c$. In Appendix~\ref{app:comp_qec}, we compare and contrast our classical process from those that describe the success rates of quantum error correction. In Appendix~\ref{app:relatedmodels}, we discuss quantum models that reduce to or are related to our classical stochastic process.

\section{Field theoretic analysis}\label{app:fieldtheory}

\subsection{Derivation of the field theory}
In this appendix, we provide explicit steps for the derivation of the field theory described in the main text. We follow the treatment in Refs.~\cite{cardytauber1998field} and~\cite{tauber2005review}; much of this derivation is standard, but we find it worthwhile to spell out relevant details. The main new contribution in this appendix is our treatment of the ``only annihilate with the nearest particle" constraint on long-range annihilation, for which we adapt results of Ref.~\cite{wijland2000hardcore}. The results of  Ref.~\cite{wijland2000hardcore} were originally intended to realize hard-core constraints, and we use them both for their original purpose and in realizing our annihilation constraint.

The starting point is a master equation that describes the reaction-diffusion system under consideration; in particular, we consider diffusion, branching and long-range annihilation processes with rates denoted by $D$, $B$ and $A$ respectively. We will define the master equation in $d$ spatial dimensions on a hypercubic lattice for concreteness. For the stochastic process in the main text, $D_0=\frac{1}{2d}(1-\pann)(1-\pbr)$, $B_0=(1-\pann)\pbr$ and $A_0=\pann$.

We define an occupation vector $\mathbf{n}$ where $n_i$ is the number of particles at site $i$. We also define unit vectors $\mathbf{e}_i$ for ease in describing changes in $\mathbf{n}$. As an example, if the number of particles on site $i$ increases by $1$, then $\mathbf{n} \to \mathbf{n} + \mathbf{e}_i$. Sometimes it is useful to directly refer to the position vector of site $i$, which we will denote as $\mathbf{r}_i$, such as in the power-law decay of the annihilation rate $1/|\mathbf{r}_i - \mathbf{r}_j|^\kappa$.

When indexing sums, we will use the notation $\sum_{j \sim i}$ to mean the sum over the set of $j$ such that $j$ is a nearest neighbor to $i$. $\sum_{j,k \sim i}$ will be the sum over the set of $j$ and $k$ such that $j \neq k$ and $j$ and $k$ are both nearest neighbors of $i$. $\sum_{j \neq i }$ will denote the sum over the set of $j$ with $j \neq i$.

The function $P(\mathbf{m};t)$ will denote the probability that the state of the system is represented by the occupation vector $\mathbf{m}$ at time $t$. The master equation for the reaction-diffusion process can be decomposed into parts representing diffusion, branching, and annihilation
\begin{align}\label{master_eq_1}
    \partial_t P(\mathbf{n};t)=P_{\textrm{diff}}+P_{\textrm{branch}}+P_{\textrm{ann}}
\end{align}
with
\begin{align}\label{master_eq_2}
    \nonumber P_{\textrm{diff}}:=&D_0\left[\sum_i \sum_{j \sim i} \left[(n_i+1)P(\mathbf{n} +\mathbf{e}_i - \mathbf{e}_j;t )\delta_{n_j,1}-n_iP(\mathbf{n};t)\delta_{n_j,0}\right]\right]\\
     \nonumber P_{\textrm{branch}} :=&
\left(B_0/ \numneigh\right)\left[\sum_i\sum_{j,k \sim i} n_i\left[P(\mathbf{n} - \mathbf{e}_j - \mathbf{e}_k; t)\delta_{n_j,1}\delta_{n_k,1}- P(\mathbf{n};t)\delta_{n_j,0}\delta_{n_k,0}\right]\right]\\
    P_{\textrm{ann}}:=& A_0 \sum_{i} \sum_{j \neq i} \frac{1}{|\mathbf{r}_i-\mathbf{r}_j|^\kappa}\left[(n_i+1)(n_j+1)P(\mathbf{n}+\mathbf{e}_i + \mathbf{e}_j; t)-n_i n_jP(\mathbf{n};t)\right]\Bigg\{\prod_{k: |\mathbf{r}_k - \mathbf{r}_i|<|\mathbf{r}_j - \mathbf{r}_i|}\delta_{n_{k},0}\Bigg\}
\end{align}
We briefly explain the above construction below. Each of these expressions $P_{\textrm{diff}}$,  $P_{\textrm{branch}}$ and $P_{\textrm{ann}}$ is the difference of two terms corresponding to the rate that the configuration $\mathbf{n}$ is entered minus the rate that the configuration $\mathbf{n}$ is left. The coefficients like $(n_i+1)$ or $n_i$ and the like are combinatorial factors that account for choosing arbitrarily any of the particles at that site to diffuse, branch or annihilate.  The diffusion and branching rates are normalized by $2d$ and $\numneigh$ to account for the fact that diffusion can cause hopping of the particle $i$ to any of its $2d$ neighbors and branching can birth particles on any of the $\numneigh$ pairs of sites neighboring $i$.

The Kronecker deltas multiplying the rates in $P_{\textrm{diff}}$ and $P_{\textrm{branch}}$ implement a \textit{hard-core} constraint for the particles, restricting to at most one particle at a site. In particular, an initial configuration that respects this constraint will continue to respect this constraint under the dynamics. We also note that we do not expect changes in universality from loosening the hard-core constraint. (More precisely, we don't expect changes as long as the on-site annihilation rate is $<\infty$. See Ref.~\cite{zhong1995onsite} for a brief discussion of certain processes without hard-core constraints always being absorbing when the onsite annihilation is infinite.)

The product of Kronecker deltas in the curly brackets in $P_{\textrm{ann}}$ implements the \textit{nearest-particle} constraint, ensuring that there are no particles closer to $i$ than $j$ when annihilating the pair; this is the factor that gives rise to the exponential term in the field theory. As noted in the main text, we \textit{do} expect this term to affect the physics, particularly at small $\kappa$, since it is responsible for regulating the long-range annihilation.

The next step in deriving the field theory in the Doi-Peliti formalism~\cite{doi1976path, peliti1985path} is to write an equivalent bosonic Hamiltonian whose \textit{imaginary-time} quantum dynamics will \textit{exactly} correspond to the real time dynamics described by the stochastic master equation. Occupation numbers $n_i$ can be interpreted as $n_i$ many bosonic particles at site $i$, and the corresponding state vector $\mathbf{n}$ can be represented as an appropriate Fock state. We introduce bosonic operators $\b_i, \b_i^\dagger$ obeying the usual algebra $[\b_i,\b_j]=[\b_i^\dagger, \b_j^\dagger]=0$ and $[\b_i, \b_j^\dagger]=\delta_{ij}$. We introduce the states $\ket{\mathbf{n}}:= \otimes_i (\b^\dagger _i)^{n_i}\ket{0}$ (this choice of normalization is standard and aids in the calculations) with the action $ \b\ket{n}=n\ket{n-1}$ and $ \b^\dagger\ket{n}=\ket{n+1}$. The occupation vector $\mathbf{n}$ can then be written as a state vector $\ket{\mathbf{n}}:= \otimes_i (\b^\dagger _i)^{n_i}\ket{0}$. 

We can then define a \textit{wavefunction} corresponding to the vector of probabilities 
\begin{equation}
\ket{\Psi(t)}:= \sum_{\mathbf{n}} P(\mathbf{n};t)\ket{\mathbf{n}}
\end{equation}
Note that this state is not normalized according to $\langle \Psi(t) | \Psi(t)\rangle = 1$; it is instead normalized according to the rule that the probabilities sum to one, which corresponds to the normalization
\begin{equation}
    \langle 0 | e^{\sum_i \hat{b}_i} |\Psi(t) \rangle = 1
\end{equation}
Expectation values in the stochastic process then become matrix elements of operators in the Hilbert space
\begin{equation}\label{eq:expectationmapping}
    \mathbb{E}[f({\bm{n}(t)})] =  \langle 0 | e^{\sum_i \hat{b}_i} f({\hat{\bm{n}}}) |\Psi(t) \rangle
\end{equation}
The time evolution of $|\Psi(t) \rangle$ takes the form of Schrodinger equation in imaginary time $t$ of the form
\begin{equation}
  -  \frac{\partial}{\partial t}\ket{\Psi(t)}= \ham \ket{\Psi(t)}
\end{equation}
for bosonic $\ham$ computed below. We emphasize again that the imaginary time dynamics with a quantum Hamiltonian actually corresponds to the real time dynamics of the classical stochastic process. 

For the master equation in Eqs. \ref{master_eq_1} and \ref{master_eq_2}, the corresponding quantum Hamiltonian is 
\begin{equation}
    \ham = \ham_\textrm{diff}+\ham_\textrm{branch}+\ham_{\textrm{ann}}
\end{equation}
with
\begin{align}\label{ham_bosonic}
   \nonumber  \ham_{\textrm{diff}} :=& -D_0\sum_{i}\sum_{j \sim i}(\b^\dagger_j\b_i-\b_i^\dagger \b_i)\delta_{\hat{n}_j,0}\\
   \nonumber \ham_{\textrm{branch}}:=&-\left(B_0/\numneigh\right)\sum_i\sum_{j,k\sim i}\b_i^\dagger\b_i\left(\b_j^\dagger\b_k^\dagger-1\right)\delta_{\hat{n}_j,0} \delta_{\hat{n}_k,0} \\
    \ham_{\textrm{ann}}:=&-A_0\sum_{i} \sum_{j \neq i} \frac{1}{|\mathbf{r}_i-\mathbf{r}_j|^\kappa} \left[\b_i\b_j-\b_i^\dagger\b_i\b_j^\dagger\b_j\right]\Bigg\{\Motimes_{k: |\mathbf{r}_k - \mathbf{r}_i|<|\mathbf{r}_j - \mathbf{r}_i|}\delta_{\hat{n}_{k},0}\Bigg\}.
\end{align}
    
The bosonic Hamiltonians are intuitive---diffusion manifests as hopping from site $i$ to $j$, branching manifests by creation operators given a particle is present at site $i$, and annihilation manifests as annihilation operators at site $i$ and $j$. The branching and annihilating operators are manifestly \textit{non-Hermitian}, implying the irreversibility of the processes. The second term in each Hamiltonian (i.e. the bosonic density operators acting as onsite energies or density-density interactions) follow from the rates of leaving configurations in the master equation; these are effectively enforcing conservation of probability. The exclusion constraints can be explicitly represented in terms of the operator $\hat{n}$ by $\delta_{\hat{n}_i,m}=\int_{-\pi}^\pi (\textrm{d}u/2\pi) e^{\textrm{i} u (\hat{n}_i-m)}$~\cite{wijland2000hardcore}.

Note that the equality of statistical expectation values with matrix elements given in Eq.~\ref{eq:expectationmapping} can be re-written as 
\begin{equation}\label{eq:expectationmapping2}
    \mathbb{E}[f({\bm{n}(t)})] =  \langle 0 | e^{\sum_i \hat{b}_i} f({\hat{\bm{n}}}) e^{-\hat{H}t}|\Psi(0) \rangle
\end{equation}
By inserting resolutions of the identity made of coherent states, we can transform this expression into a coherent state path integral. There are subtleties, especially with time ordering and boundary conditions, and we direct the interested reader to section 3.3 of Ref.~\cite{tauber2005review} and references therein for a detailed discussion of the coherent state path integral in this context.

We consider coherent states labeled by a vector of complex numbers $\bm{\phi}$ and denoted by $\ket{\bm{\phi}}$. These states are defined as eigenstates of the annihilation operator, $\hat{b}_i \ket{\bm{\phi}} = \phi_i \ket{\bm{\phi}}$, and constructed as $\ket{\bm{\phi}}=e^{\sum_i\phi_i\b_i^\dagger}\ket{0}$. For this choice of normalization, the resolution of the identity is 
\begin{equation}
\begin{split}\label{eq:resolution}
\unit=\int \left(\prod_j(\textrm{d} \phi_j\textrm{d}\conjphi_j/2\pi i)\right)e^{-\sum_j\phi_j\conjphi_j}\ket{\bm{\phi}}\bra{\bm{\phi}}.
\end{split}
\end{equation}
The integral is to be interpreted as follows. We can decompose $\bm{\phi}$ into its real and imaginary parts $\bm{\phi}= \bm{R} + i \bm{I}$. The integral in Eq.~\ref{eq:resolution} becomes 
\begin{equation}
\begin{split}
\unit=\int \left(\prod_j(\textrm{d} R_j\textrm{d}I_j/\pi)\right)e^{-\sum_j(R_j^2 + I_j^2)}\ket{\bm{R} + i \bm{I}}\bra{\bm{R} + i \bm{I}}.
\end{split}
\end{equation}
where all integrals run from $-\infty$ to $\infty$.

As noted, we can compute a path integral by inserting resolutions of the identity into time-slices of Eq.~\ref{eq:expectationmapping2}. $e^{-\hat{H} t}$ will become a weighting $e^{-S}$ with $S$ a functional of $\bm{\phi}(t)$ and $e^{-S}$ understood perturbatively~\cite{tauber2005review}. It is nice to write $H$ in a normal-ordered form, as this allows the $b$'s of the Hamiltonian to become the $\phi$'s of one time slice and the $b^\dagger$'s to be the $\phi^*$'s of the next time slice. Besides a time derivative term and boundary terms, $S$ can then be written down by computing $\frac{\langle \bm{\phi}|H| \bm{\phi} \rangle}{\langle \bm{\phi}|\bm{\phi} \rangle}$:
\begin{equation}\label{eq:actionfromexpectation}
    S[\bm{\phi}] = \text{boundary terms} + \int_0^{t_f} dt \left[ \frac{\langle \bm{\phi}|H| \bm{\phi} \rangle}{\langle \bm{\phi}|\bm{\phi} \rangle} + \sum_i \conjphi_i \partial_t \phi_i \right]
\end{equation}
where $\bm{\phi}$ is a function of the integration variable $t$. It is nice to use $t$ as the dummy integration variable; when calculating observables at time $t$, $t_f$ will then be $t$ in a slight abuse of notation. As noted, we direct the interested reader to section 3.3 of Ref.~\cite{tauber2005review} and references therein for more details.

The only non-trivial normal-ordering is of the exclusion constraint operators $\delta_{\hat{n}_i,0}$ and $\delta_{\hat{n}_i,1}$. The normal ordering can be computed~\cite{wijland2000hardcore} using the integral representation mentioned before. For our purposes, we only need the following identity, which holds for integers $m\geq 0$, 
\begin{equation}
    \begin{split}
        \frac{\mel{\phi}{\delta_{\hat{n},m}}{\phi}}{\bra{\phi} \ket{\phi}}&=\frac{1}{m!}(\conjphi\phi)^me^{-\conjphi\phi}
    \end{split}
\end{equation}
from which related identities can be derived straightforwardly, including $\frac{\mel{\phi}{b^\dagger \delta_{\hat{n},m}}{\phi}}{\bra{\phi} \ket{\phi}} = \phi^*  \frac{\mel{\phi}{\delta_{\hat{n},m}}{\phi}}{\bra{\phi} \ket{\phi}}$ or $\frac{\mel{\phi}{b \delta_{\hat{n},m}}{\phi}}{\bra{\phi} \ket{\phi}} = \frac{\mel{\phi}{ \delta_{\hat{n},m-1} b}{\phi}}{\bra{\phi} \ket{\phi}} = \phi  \frac{\mel{\phi}{\delta_{\hat{n},m-1}}{\phi}}{\bra{\phi} \ket{\phi}}$. Note that $\frac{\mel{\phi}{\delta_{\hat{n},m}}{\phi}}{\bra{\phi} \ket{\phi}}=0$ for $m<0$. 

For diffusion:
\begin{equation}\label{eq:coherentdiff}
\begin{split}
    \frac{\langle \bm{\phi}|H_{\rm diff}| \bm{\phi} \rangle}{\langle \bm{\phi}|\bm{\phi} \rangle} &= -D_0\sum_i \sum_{j \sim i} (\conjphi_j \phi_i - \conjphi_i \phi_i)e^{-\conjphi_j \phi_j} 
    \\ &= -D_0\sum_i \sum_{j \sim i} (\conjphi_j \phi_i - \conjphi_i \phi_i) + \text{ subleading}
\end{split}
\end{equation}
Near the critical point and in the absorbing phase, we can Taylor-expand the exponential. The contributions labeled subleading in Eq.~\ref{eq:coherentdiff} will ultimately be subleading in the RG sense after taking the continuum limit, as they will contain too many fields and derivatives. 
Similarly, for annihilation:
\begin{equation}\label{eq:coherentann}
\begin{split}
\frac{\langle \bm{\phi}|H_{\rm ann}| \bm{\phi} \rangle}{\langle \bm{\phi}|\bm{\phi} \rangle}=&-A_0\sum_{i} \sum_{j \neq i} \frac{1}{|\mathbf{r}_i-\mathbf{r}_j|^\kappa} \left[\phi_i\phi_j-\conjphi_i \phi_i\conjphi_j\phi_j\right] e^{-\sum_{k: |\mathbf{r}_k - \mathbf{r}_i|<|\mathbf{r}_j - \mathbf{r}_i|} \conjphi_{k}\phi_{k}}
\end{split}
\end{equation}
Note that we do not attempt to Taylor expand the argument of the exponential. The reason is that although each individual $\barphi_{k}\phi_{k}$ may be small close to the transition and in the absorbing phase, their sum might be considerable if sites $i$ and $j$ are far apart. We will indeed find that this is the case when we consider the saddlepoint equations.
For branching:
\begin{equation}\label{eq:coherentbranch}
\begin{split}
    \frac{\langle \bm{\phi}|H_{\rm branch}| \bm{\phi} \rangle}{\langle \bm{\phi}|\bm{\phi} \rangle} &= -\left(B_0/\numneigh\right)\sum_i\sum_{j,k\sim i}\conjphi_i \phi_i (\conjphi_j \conjphi_k-1) e^{- \conjphi_j \phi_j - \conjphi_j \phi_k}
    \\ &= -\left(B_0/\numneigh\right)\sum_i\sum_{j,k\sim i}\conjphi_i \phi_i ( (\conjphi_i + (\conjphi_j - \conjphi_i)) (\conjphi_i + (\conjphi_k - \conjphi_i))-1) e^{ - \conjphi_j \phi_j - \conjphi_j \phi_k}
    \\&= -\left(B_0/\numneigh\right)\sum_i\sum_{j,k\sim i}\conjphi_i \phi_i ( {\conjphi_i}^2-1)  + \text{ subleading} 
    \\&= -B_0 \sum_i \conjphi_i \phi_i ( {\conjphi_i}^2-1) + \text{ subleading}
\end{split}
\end{equation}
For branching, almost all of the terms labeled subleading are indeed subleading in the RG sense, as they contain products of many fields and/or derivatives. However, one of the terms coming from the linear order of the Taylor expansion of the exponential is $\conjphi_i \phi_i\conjphi_j\phi_j$ with $j$ and $i$ nearest neighbors, which is comparable to the analogous term in the annihilation process. We neglect this contribution for simplicity, as we view it as effectively already contained in the annihilation process. Additionally, when considering the saddlepoint equations, such a contribution will enter at $O(n^2)$.

As noted above, these expectation values will ultimately yield the action $S$ of the field theory, up to a time derivative term and boundary terms. $f(\hat{n})$ will give rise to fields multiplying $e^{-S}$ in the path integral. Writing $f(\hat{\bm{n}})$ explicitly in terms of $\hat{b}$ and $\hat{b}^\dagger$ in a normal-ordered form, the $\hat{b}^\dagger$ acting to the left on $\langle 0 | e^{\sum_i \hat{b}_i}$ will be just $1$, leaving only products of $\hat{b}_i$ terms. After inserting resolutions of the identity to construct the path integral, $\hat{b}_i \to \phi_i(t)$. $|\Phi(0)\rangle$ and $\langle 0 | e^{\sum_i \hat{b}_i}$ will give boundary conditions of the path integral; the $\langle 0 | e^{\sum_i \hat{b}_i}$ boundary condition is especially straightforward, as this is in fact the coherent state $\langle \bm{1}|$.

For example, following this procedure, $\mathbb{E}[n_i(t)] = \langle \phi_i(t) \rangle$ and $\mathbb{E}[(n_i(t))^2] = \langle (\phi_i(t))^2 + \phi_i(t) \rangle$, where $\langle \rangle$ means expectation value with respect to the above-mentioned action and boundary conditions. If we refrain from Taylor expanding the exponentials implementing the hard-core constraint and then dropping subleading terms, the coherent state path integral at this stage will still reproduce the statistical expectation values. However, we indeed drop terms and will subsequently take a continuous space limit. These approximations will maintain universality but will no longer allow for exactly reproducing statistical expectation values.

We will focus on the action in the following, and we will not further discuss boundary conditions. Taking the continuous space limit of Eq.~\ref{eq:actionfromexpectation} using Eqs.~\ref{eq:coherentdiff}, \ref{eq:coherentann}, and \ref{eq:coherentbranch} (with the choice $\phi_i(t) \to a^d \phi(\mathbf{x},t)$ and $\conjphi_i(t) \to \barphi(\mathbf{x},t)$ for lattice spacing $a$), we obtain the action
\begin{equation}\label{action}
    S:= \int_{t_1}^{t_2}\textrm{d} t\:\left[ \int \textrm{d}^d{\mathbf{x}}\left[\barphi\partial_t\phi+\mathcal{H}_{\textrm{local}}(\barphi;\phi)\right]+\int \textrm{d}^d \mathbf{x} \textrm{d}^d\mathbf{y} \: \mathcal{H}_{\textrm{non-local}}(\barphi;\phi)\right]
\end{equation}

\begin{align}\label{hamiltonian_density}
    \nonumber \mathcal{H}_{\textrm{local}}&:=-D \barphi\nabla^2\phi +B(1-{\barphi}^2)\barphi\phi \\
    \mathcal{H}_{\textrm{non-local}}&:=- A_{\kappa}(|\mathbf{x}-\mathbf{y}|)\left(1-\barphi(\mathbf{x},t)\barphi(\mathbf{y},t)\right)\phi(\mathbf{x},t)\phi(\mathbf{y},t) \: \textrm{exp}\left(-\int_{r=0}^{r=|\mathbf{x}-\mathbf{y}|} \textrm{d}^d \mathbf{r}\:   \barphi(\mathbf{x}+\mathbf{r},t)\phi(\mathbf{x}+\mathbf{r},t) \right)
\end{align} 
where $A_\kappa(|\mathbf{x}-\mathbf{y}|):=A/|\mathbf{x}-\mathbf{y}|^\kappa$  Eqs. \ref{action} and \ref{hamiltonian_density} describe the field theory referenced in the main text. $D,B,A$ are equal to $D_0, B_0, A_0$ up to factors of the lattice spacing. We will generically take $t_1 = 0$ and $t_2 = t$.

Before concluding this subsection, we highlight certain salient features of this field theory. (1) We have obtained a non-relativistic field theory (since the kinetic term is given by $\barphi(\partial_t-D\nabla^2)\phi$) for a complex scalar field $\phi$. (2) The branching term contributes to a mass $B$ for the fields and can be appended to the Gaussian part of the action. The branching term also contributes an interaction term of the form ${\barphi}^3\phi$; the annihilation term completely contributes to interaction terms, of which key to the annihilation is $\phi(\mathbf{x},t)\phi(\mathbf{y},t)$ while the other term is a non-local density-density interaction, and both of these terms are weighted by the nearest-particle exponential factor. We also note that the field theory is non-local only in space, and remains local in time. (3) The lack of particle number conservation manifests in the non-Hermitian nature of the field theory and lack of $U(1)$ symmetry. The explicit parity conservation of the dynamics corresponds to $\mathbb{Z}_2$ symmetry of $\phi\mapsto -\phi$ and $\barphi\mapsto -\barphi$. (4) The scaling dimensions of the coupling constants can be straightforwardly computed using $[\textrm{length}] =-1$ and $[\textrm{time}]=-2$ so that $[\textrm{momentum}]=1$; under renormalization, there will be anomalous dimensions ascribed to fluctuation effects over mean-field level.

\subsection{Euler-Lagrange minimization}\label{app:EL_min}
In this subsection, we provide additional details about the mean-field analysis mentioned in the main text. The mean-field rate equation is obtained by Euler-Lagrange minimization of the action derived in the previous subsection for the fields $\phi$ and $\barphi$. The Euler-Lagrange equations we obtain are as follows,
\begin{subequations}
    \begin{align}
         \nonumber &\left[\partial_t-D \nabla^2+B-3B{\barphi}^2(\mathbf{x},t)+\int \textrm{d}^d\mathbf{z}\left[A_\kappa(|\mathbf{x}-\mathbf{z}|)\barphi(\mathbf{z},t)\phi(\mathbf{z},t)(\expterm( \mathbf{z}, \mathbf{x};t )+\expterm( \mathbf{x}, \mathbf{z};t ))\right]\right]\phi(\mathbf{x},t)-\\
\label{EL_1} &\left[\int \textrm{d}^d\mathbf{z}_1\textrm{d}^d\mathbf{z}_2\: A_{\kappa}(|\mathbf{z}_1-\mathbf{z}_2|)\left(1-\barphi(\mathbf{z}_1,t)\barphi(\mathbf{z}_2,t)\right)\phi(\mathbf{z}_1,t)\phi(\mathbf{z}_2,t)\expterm( \mathbf{z}_1, \mathbf{z}_2;t )\bm{1}_{0\leq |\mathbf{z}_1-\mathbf{x}|\leq |\mathbf{z}_1-\mathbf{z}_2|}\right]\phi(\mathbf{x},t)=0\\
   \nonumber  &\left[-(\partial_t +D\nabla^2)+B-B{\barphi}^2(\mathbf{x},t)-\int \textrm{d}^d\mathbf{z}\: A_{\kappa}(|\mathbf{x}-\mathbf{z}|)(1-\barphi(\mathbf{x},t)\barphi(\mathbf{z},t))\phi(\mathbf{z},t)(\expterm( \mathbf{z}, \mathbf{x};t )+\expterm( \mathbf{x}, \mathbf{z};t ))\right]\barphi(\mathbf{x},t)+\\
     \: 
    \label{EL_2} &\left[\int \textrm{d}^d\mathbf{z}_1\textrm{d}^d\mathbf{z}_2\: A_{\kappa}(|\mathbf{z}_1-\mathbf{z}_2|)\left(1-\barphi(\mathbf{z}_1,t)\barphi(\mathbf{z}_2,t)\right)\phi(\mathbf{z}_1,t)\phi(\mathbf{z}_2,t)\expterm( \mathbf{z}_1, \mathbf{z}_2;t )\bm{1}_{0\leq |\mathbf{z}_1-\mathbf{x}| \leq |\mathbf{z}_1-\mathbf{z}_2|}\right]\barphi(\mathbf{x},t)=0
\end{align}
\end{subequations}
where $\expterm(\mathbf{x},\mathbf{y};t):=\textrm{exp}\left(-\int_{r=0}^{r=|\mathbf{x}-\mathbf{y}|} \textrm{d}^d \mathbf{r}\:   \barphi(\mathbf{x}+\mathbf{r},t)\phi(\mathbf{x}+\mathbf{r},t) \right)$ is a shorthand notation for the exponential factor implementing the nearest-particle constraint and $\mathbf{1}_{x \in \textrm{support}}$ is the indicator function on the real line. While the Euler-Lagrange equations look complicated, a remarkable simplification is obtained by noting that $\barphi=1$ is a solution of Eq. \ref{EL_2} (this occurs generically~\cite{tauber2005review}). Using this solution in Eq. \ref{EL_1}, and identifying the mean-field $\phi$ as the local density $n$, we obtain the rate equation,
\begin{equation}
\dot{n}(\mathbf{x},t) = D \nabla^2 n(\mathbf{x},t) + 2B n(\mathbf{x},t) - A n(\mathbf{x},t)   \int d^d \mathbf{y} \frac{n(\mathbf{y},t)}{|\mathbf{x}-\mathbf{y}|^\lrpow} \left( e^{-\int_{r=0}^{|\mathbf{x}-\mathbf{y}|} d^dr n(\mathbf{x}+\mathbf{r}, t)} + e^{-\int_{r=0}^{|\mathbf{x}-\mathbf{y}|} d^dr n(\mathbf{y}+\mathbf{r}, t)} \right)
\end{equation}
To get to this rate equation, we made several approximations in the field theory; explicitly including the exponentials from the hard-core constraints (particularly on the branching term, see discussion below Eq.~\ref{eq:coherentbranch}) will give corrections of size at most $O(n^2)$.

We will further assume a uniform initial condition; this equation is translationally invariant, and so $n(x,t)$ will then be independent of $x$ at all times. Introducing $\mathbf{z} = \mathbf{x}-\mathbf{y}$ and the volume of a $d$-dimensional unit ball $V_d$, the dynamics further simplify to
\begin{equation}
\begin{split}
\dot{n}(t) = 2B n(t) - 2 A n(t)^2 \int d^d z \frac{1}{z^\lrpow} e^{-V_d z^d n(t)}
\end{split}
\end{equation}
For $\lrpow \leq d$, the integral needs the infrared-regulator provided by the decaying exponential, which forces the resulting integral to depend strongly on $n$. When $\lrpow > d$, there is no longer an infrared divergence, but there is an ultraviolet divergence; this small-$z$ divergence in the integral is regulated by the lattice spacing, and so $\int d^d z \frac{1}{z^\lrpow} e^{-V_d z^d \rho} \sim \text{const} + O(n(t))$. Together, these give (with non-universal constant $c$ depending on $d,\lrpow, a$)
\begin{equation}\label{eq:meanfieldrate}
\dot{n}(t) \sim 2B n(t) - cA \begin{cases} n(t)^{1+\frac{\lrpow}{d}} & \lrpow \leq d 
    \\ n(t)^2 \log(\frac{1}{n(t)}) & \lrpow = d
    \\  n(t)^2 & \lrpow > d
    \end{cases} 
\end{equation}
To reiterate, at the level of mean-field theory, the exponential in the annihilation term prevents the infrared divergence of the long-range annihilation double-integral at small $\lrpow$ by introducing a density-dependent cutoff length.

Note that mean field theory only holds in sufficiently high dimensions, if at all. Indeed, our mean field theory does not predict an absorbing phase for any $\lrpow > 0$ at any positive branching $B$; a positive coefficient on the dominant term $n(t)$ forces a non-vanishing steady state $n(t) \sim \text{const}$. This was argued (in the case of $\kappa = \infty$) to be an artifact of the infinite-dimensional nature of mean field theory~\cite{cardytauber1998field}; a special upper critical dimension $d_c$ was identified below which the absorbing phase is indeed stable. For $\lrpow=\infty$, $d_c$ was found to be $4/3$ at one-loop order~\cite{cardytauber1998field}; on the other hand, our mean-field equations at $\lrpow=0$ show a transition at nonzero branching and are hence consistent with $d_c = \infty$. We thus anticipate that $d_c$ will be a function of $\kappa$. 

Despite the omission of the absorbing phase at nonzero branching for $\kappa>0$, the mean-field predictions for the absorbing phase -- i.e. solving Eq.~\ref{eq:meanfieldrate} when $B=0$ -- match that of our scaling arguments when $d \geq 2$ in Eq.~4 of the main text (up to logarithmic corrections when $\lrpow = d$). We view this agreement in high dimensions as signaling that field theory is an appropriate tool for understanding properties of the absorbing phase and the transition. However, to study low dimensions and go beyond mean-field theory, it is necessary to consider renormalization. 
We note that both perturbative RG (once appropriate counterterms are identified for the exponential factor) in the style of Cardy and Täuber~\cite{cardytauber1998field}, and non-perturbative RG in the style of section 3.4.2 of the review~\cite{dupuis2021nonperturbative} appear promising. We welcome further studies of this unusual field theory, which is interesting in its own right.

\section{Subleading corrections in the diffusion-dominated absorbing phase}\label{app:subleading}

In the main text, we discuss the behavior of absorbing phase decay exponent $\alpha$ ($n(t) \sim t^{-\alpha}$) as a function of $\lrpow$. One of our predictions in $d=1$ is that for every $\lrpow \geq 2$, $\decaypow = \frac{1}{2}$. This is borne out in our numerics except for a small discrepancy that is most pronounced when $\lrpow \approx 2.6$. In this brief appendix, we discuss this discrepancy in more detail, and why we believe it  comes from finite-time transients in the decay.

There are several ways to describe the density decay $n(t) \sim t^{-1/2}$ of a diffusion-limited pure annihilation process in $d=1$. As noted in chapter 13 of Ref.~\cite{shlomo2000textbook}, one method is to use an effective rate equation of the form $\dot{n}(t) \sim -n(t)^3$ which, despite being deterministic and having no diffusion term, yields the correct exponent of the decay. We will combine this effective rate equation with the contribution $n(t)^{\kappa+1}$ to the decay from the long-range annihilation, giving 
\begin{equation}\label{eq:effratesubleading}
    \dot{n}(t) \sim -c_d n(t)^3 - c_l n(t)^{\kappa + 1}
\end{equation}
where $c_d$ and $c_l$ are $\kappa$-dependent constants multiplying the terms corresponding to diffusion and long-range annihilation.
We emphasize that this effective rate equation is heuristic and likely will not capture the quantitative forms of the subleading asymptotics, but we expect this equation will successfully describe their qualitative features.

For $\kappa \geq 2$, the long-range term $n(t)^{\kappa + 1}$ is subleading relative to the leading diffusive term $n(t)^3$. It is useful to consider different regimes of $\kappa \geq 2$.

When $\kappa$ is sufficiently large, the long-range term rapidly vanishes relative to the diffusive term and will not have an effect on the dynamics outside of a short-lived transient. This would explain why $\kappa \geq 3$ does not show a noticeable discrepancy besides early times (the $\kappa=3$ data are shown in the inset of Fig.~2). 

When $\kappa$ is close to $2$, $\kappa = 2 + \epsilon$ with $\epsilon$ very small, the long-range term is comparable in magnitude to the diffusive term for a time that diverges as $\epsilon \to 0$. However, despite being long-lived, the subleading term is so similar to $n(t)^3$ for small $\epsilon$ that it then cannot noticeably change the functional form of the decay. This would explain why we do not see a discrepancy for $\kappa=2$.

To summarize, it is only for $\kappa$ close but not too close to $2$ that the subleading term will lead to a long-lived transient that is also sufficiently different from the asymptotic decay to be noticeable. This explains why the discrepancy goes to zero at $\kappa=2$ and $\kappa \geq 3$; the subleading term needs to strike the right balance. 

Eq.~\ref{eq:effratesubleading} would also predict subleading transients for $\kappa < 2$. However, it is likely that the correct effective rate equation will have a rather suppressed diffusive term $n(t)^3$. The reason is that in the long-range-dominated regime of $\kappa < 2$, long-range annihilation will with very high probability destroy far-away pairs of particles before they can diffuse near each other, removing much of the diffusive contribution to annihilation. This should be contrasted to what happens in the diffusion-limited regime. Although particles coming close to one another from diffusion is what sets the asymptotic decay rate, long-range annihilation events will still occur. They will be subleading, rather than both subleading and suppressed.

\section{Dynamical exponent}\label{app:dynamical}

\begin{figure}[t]
\includegraphics[width=.6\linewidth]{fig4.png}
\caption{Dynamical exponent $z_c$ at the transition. Between $2000$ - $5000$ samples were averaged to construct $n(t)$ at each $\kappa$ and system size $L$ probed. For each $\kappa$, $p$ was set to the corresponding $\pc$ extracted from Fig.~3 of the main text. At a given $\kappa$, we fit $t^{\alpha_c} n(t)$ to $e^{-t/\tau(L)}$ for each $L$, and then extract $z$ from $\tau(L) \sim L^z$. Standard deviations are estimated with 100 bootstrap resamples of $n(t)$. The values used for the dotted lines for $z_{\rm PC}$ and $z_{\rm DP}$ are taken respectively from~\cite{grassberger2025precision} ($1.74104(4)$) and~\cite{jensen1999series} ($1.580745(10)$). Inset shows example collapse for $\lrpow = 2$.}
\label{fig:transition_z}
\end{figure}

In the main text, we numerically estimated the decay exponent $\alpha_c$ at the critical point between the absorbing and active phases. The behavior of the exponent was consistent with $\alpha_c = \alpha_{\rm DP}$ at $\kappa=0$, $\alpha_c = \alpha_{\rm PC}$ for $\kappa > 2$, and interpolating between the two for $0<\kappa<2$. In this appendix, we investigate another independent exponent and show that it enjoys similar behavior. 

In Fig.~\ref{fig:transition_z}, we consider the dynamical exponent $z_c$, which we extract through the time-dependence of the decay of particle density at criticality.  In an infinite system, the decay is controlled by the power law $t^{-\alpha_c}$, but in finite-size systems at sufficiently late times, the density decays much more rapidly on timescales set by $L^{z_c}$. In particular, the scaling hypothesis predicts $n(t) \sim t^{-\alpha_c} \Phi(t/L^{z_c})$ for some scaling function $\Phi$; empirically, $\Phi(x)\sim e^{-c x}$. Thus, by fitting $t^{\alpha_c} n(t)$ to an exponential decay in time for several values of $L$, one can extract $z_c$. This method of extracting $z_c$ gives rise to good collapse over several orders of magnitude of $t^{\alpha_c} n(t)$. We note that although the difference between $z_{\rm PC}$ and $z_{\rm DP}$ is relatively small, we can still distinguish the exponents, and we again see a clear interpolating regime. 

Our standard-deviation errorbars only reflect the uncertainty in our $n(t)$ runs, which we take for values of $L$ between $L=40$ and $L=800$ and run for times $\approx 10 L^{z_c}$. The error bars do not reflect the (small) additional uncertainty coming from the fact that the $p_c$ and $\alpha_c$ we use are themselves uncertain.

\section{Differences from quantum error correction}\label{app:comp_qec}
A canonical task in quantum information theory is to prepare a desired many-body entangled state, such as a topologically ordered ground state of the 2D toric code~\cite{kitaev2003fault} or a Schrodinger-cat ground state of the 1D Ising model. A typical ``low-entropy" configuration around the desired state often looks like a dilute gas of point-like excitations (such as anyons in the 2D toric code or domain-walls in a 1D Ising model). In the simplest cases, these excitations may only by annihilated or created in pairs, and they may be separated freely without energy penalty.   Canonical measurement-based state-preparation protocols make measurements everywhere in the system to deduce the locations of excitations, and apply non-local ``string-like" recovery operations to remove excitations in pairs~\cite{BriegelRaussendorf, RBH, PiroliCirac}. This process can be mapped to a classical stochastic process describing the dynamics of excitations. However, crucially,  the long-range nature of feedback means that pairs of particles can effectively instantaneously annihilate over long distances. 

While our ``attempt annihilation only with the closest particle" rule was inspired by quantum error correction, we note some key differences between our absorbing state transition and error correction thresholds in the literature. In particular, our absorbing state dynamics do not keep track of whether a ``logical error" occurred (the primary object of interest in quantum error correction). Furthermore, in quantum error correction, errors may occur everywhere, but our dynamics only allow the spawning of new defects via local particle branching. This restriction allows for the existence of an absorbing state. 

There are similarities, in that the proliferation of defects is what can induce a logical error when attempting to decode in quantum error correction, but the two problems are ultimately concerned with different quantities. 

However,  the two can be intimately tied together in the related context of heralded noise. In Ref.~\cite{chirame2024stableSPT}, classical flags are used to track information about potential error locations, and these flags can reach an absorbing state. Decoding operations are conditional on the classical flags, and the absorbing state for the classical flags halts all these decoding operations. This effectively forces the logical failure rate to be $O(1)$ in the classical flag absorbing phase, tying the logical information and absorbing state transitions together.

We discuss further relations between quantum and classical processes in Appendix~\ref{app:relatedmodels}, including other classes of quantum models that display the same universality classes as the classical models we study.
\section{Related models from random circuits with feedback}\label{app:relatedmodels}
We discuss a few ways for random quantum circuits to realize models in the same universality class as the one studied in the main text. Mapping quantum circuits to classical stochastic processes is a standard technique; for a review, see~\cite{fisher2023review} and references therein. Appendix I of Ref.~\cite{odea2024entanglement} is also useful for a review of mappings in cases involving measurement and feedback, and we summarize the main idea here. Averaging over block-Haar gates gives rise to a stochastic process described by block-diagonal stochastic matrices satisfying detailed balance, while feedback based on measurement outcomes can violate detailed balance and induce interesting non-equilibrium behavior like absorbing state transitions. 

The average over random circuit realizations can be realized as a channel acting on the averaged density matrix of the system. As noted in Ref.~\cite{odea2024entanglement}, for appropriate choices of random gates and feedback, the diagonal and off-diagonal of the density matrix decouple. The diagonal of the density matrix can be viewed as a $2^L$ vector of probabilities that is time-evolved by a stochastic matrix. This time-evolution is equivalent to the stochastic evolution of a particle system on $L$ sites, allowing the dynamics of certain observables to be described by a simple classical Markov process. In the case of a direct realization of the stochastic process on the sites, the time evolution of all diagonal observables are encoded in the classical process.

We can consider cases where the particles of the stochastic process don't live directly on the sites; for example, they might correspond to domain walls living halfway between sites in a $1$d system, or anyons living on the centers of plaquettes in a $2$d system. The domain walls would then be probed through $\sigma^z_i \sigma^z_{i+1}$ measurements and the anyons through plaquette measurements. However, for simplicity, we will walk through the simpler special case where the particles of the stochastic process live directly on sites. The main difference in the examples with domain walls and anyons is a redundancy of the classical description, where a given description in terms of classical particles may correspond to a handful of different quantum density matrices. Despite this redundancy, appropriate local observables can still be extracted from the classical stochastic process, such as $\mathbb{Z}_2$-symmetric diagonal operators in the case of a $1$d system with domain walls.

We will consider a realization of our 1d process directly on a lattice of qubits. All operations on the quantum system will be in the computational basis generated by $\ket{\uparrow}$ and $\ket{\downarrow}$, while the corresponding operations on the related classical process will be described in terms of the action on bitstrings generated by $0$ and $1$ corresponding respectively to $\ket{\uparrow } \bra{\uparrow}$ and $\ket{\downarrow} \bra{\downarrow}$. That is, spin-down on a single site will ultimately correspond to a particle in the classical stochastic process occupying that site, while spin-up corresponds to an empty site.

A two-site particle number conserving block-diagonal Haar matrix $U_h$ gives rise to a hopping term in the classical process. In particular, averaging $U_h \otimes U^*_h$ gives rise to a matrix that is block-diagonal between the diagonals and off-diagonals of the density matrix. In particular, there is a block corresponding to $\ket{\uparrow \uparrow} \bra{\uparrow \uparrow},\ket{\uparrow \downarrow} \bra{\uparrow \downarrow},\ket{\downarrow \uparrow} \bra{\downarrow \uparrow},\ket{\downarrow \downarrow} \bra{\downarrow \downarrow}$, which we will identify with the bitstring basis $\{00, 01, 10, 11\}$. In this basis, the block takes the form 
\begin{equation}
    \begin{pmatrix}
        1 & 0 & 0 & 0 \\
        0 & \frac{1}{2} & \frac{1}{2} & 0 \\
        0 & \frac{1}{2} & \frac{1}{2} & 0 \\
        0 & 0 & 0 & 1
    \end{pmatrix}
\end{equation}
corresponding to unbiased hopping between $01$ and $10$. 

Similarly, a two by two block in a three-site Haar matrix between $\ket{\uparrow \downarrow \uparrow}$ and $\ket{\downarrow \downarrow \downarrow}$ gives rise to processes taking $010 \leftrightarrow 111$ and back corresponding to branching and local annihilation. Alternatively, if there are relevant non-unitary processes, one could have the processes $010 \leftrightarrow 111$ occur with unequal probabilities. In our model, we assume this to be the case, taking all of our annihilation to come from feedback conditioned on measurements rather than additionally from local unitary processes. However, we do not find significant differences between models where $010 \leftrightarrow 111$ occurs locally with equal probabilities and models where only $010 \rightarrow 111$ occurs (modulo the feedback steps in these models). 

Feedback allows for nonlocal annihilation processes. In our model, measurements are assumed to be taken everywhere so that all defect positions are known. A particle is selected at random for its dynamics, including whether or not to attempt to pair it with the closest particle to it and attempt annihilation. Alternatively, using the ingredients described above, we could have constructed models with parallel update rules, involving parallelized feedback and brickwork circuits of unitary gates to induce hopping and branching. 

Our serial model has some differences relative to the parallel models; the parallel models may be more natural, given that defects are paired all at once in minimum-weight perfect matching decoders. We chose our serial model for our numerics because we found that it has better finite-size and -time effects, making it easier to illustrate our results.

We also found that some choices of parallel feedback could induce long-range attraction between the defects, even in the ostensibly short-range limit of $\lrpow = \infty$. The idea is that if defects are always paired to minimize the sum total of the weights between pairs, there are certain processes occurring at nonzero probability that effectively induce biased random walks. For example, consider the following hypothetical process where two far-away particles both branch and then undergo the annihilation feedback step:
\begin{align*}
    &00000100000100000\\
    &00001110001110000\\
    &00000010001000000
\end{align*}
Because the feedback minimizes the sum of the distances, the two particles at the edge of the cluster of $1$s closest to the opposite cluster always pair. Though this pair won't annihilate, this pairing forces the pairing and annihilation of the remaining two pairs of adjacent particles in the clusters, leaving just those ``inner" two $1$'s at the final time step. This is effectively inducing a biased motion between the $1$s closer to each other; while there are other possible processes or orderings of processes that could occur, there is a net bias even when taking those into account. Biased walks that favor attraction between particles can be relevant for the universality class of the transition~\cite{suchanpark2020first, suchanpark2020second}. Given our interest in interpolating between a long-range directed percolation and a short-range parity conserving limit, we also chose our serial process to avoid these biased walks. However, we believe it would be interesting in future work to compare the bias induced from the above parallel process to bias induced in other ways in the literature.

In this appendix, we have outlined several ingredients in quantum circuits with measurements and feedback that give rise to models similar to the one numerically probed in the main text. While many of these models fall in the same universality class as the model in the main text, there are some that may fall outside, particularly those with biased random walks attracting particles towards one another. We note that there are further variations on measurement and feedback, including using ``flags" both for measurement outcomes and for error locations~\cite{chirame2024stableSPT}; these introduce further classical degrees of freedom which may themselves undergo absorbing state transitions.

\end{document}